\documentclass[aps,prb,showpacs,twocolumn,floats]{revtex4}
\usepackage{amssymb}
\usepackage{amsbsy}
\usepackage{amsmath}
\usepackage{epsfig}

\begin{document}

\title {Model study of dissipation in quantum phase transitions }

\author{Subhasis Sinha and Sushanta Dattagupta}

\affiliation{Indian Institute of Science Education and
Research-Kolkata, Mohanpur, Nadia 741252, India.}

\begin{abstract}
We consider a prototypical system of an infinite range transverse field Ising 
model coupled to a bosonic
bath. By integrating out the bosonic degrees, an effective anisotropic 
Heisenberg model is obtained for the spin system. The phase diagram of the 
latter is calculated as a function of coupling to the heat bath and 
the transverse magnetic field. Collective excitations at low temeratures are 
assessed 
within a spin-wave like analysis that exhibits a vanishing energy gap at 
the quantum 
critical point. We also consider another limit where the system reduces to
a generalized spin-boson model of two interacting spins. By increasing the 
coupling strength with the heat bath, the two-spin wavefunction changes from
an entangled state to a factorized state of two spins which are aligned along 
the
transverse field. We also discuss the possible realization and application 
of the model to different physical systems.

\end{abstract}


\maketitle

\date{\today}

\section {Introduction}
The twin (and apparently disjoint) topics of quantum phase transition (QPT) and quantum dissipation (QD) have seen a great upsurge of activity in recent years. The QPT is of significance in many areas of contemporary interest in the Condensed Matter, such as metal-insulator transition, quantum magnetism, 
ferroelectricity, superconductivity and in the general question of coherence and quantum computation \cite{sachdev,girvin}. On the other hand, QD is 
ubiquitously present because of environmental influences on otherwise unitary evolution of quantum systems\cite{weiss}. Dissipation, though considered a pest, needs to be understood (and tamed), in order to tackle decoherence effects in quantum many body systems. Our aim in this paper is to analyze the combined presence of these two seemingly disparate phenomena of QPT and QD via simple model systems. Our hope is to elucidate on the irreversible effects near a quantum critical point(QCP) because of dissipative interactions with the environment.

The simplest model of a QPT is an Ising system of coupled spins (which are taken to be polarized in one direction only, say the z-axis), subjected to an additional ('magnetic') field $\Gamma$ along, say the x-axis. The latter couples to the x-component of the spins which, because of its non-commutativity with the z-component, triggers quantum dynamics in the system. The resultant 'Transverse Ising Model (TIM)' is a prototype for analyzing the conflicting presence  of 'order' along z-axis induced by the Ising coupling and `disorder', caused 
by the tilting of the spins to the x-axis by $\Gamma$. The net result is the occurrence of a QCP at temperature $T=0$ in the $\Gamma - T$ phase diagram as $\Gamma$ is increased 
to a critical value $\Gamma_c$, when the system transits from a ferromagnetic to a paramagnetic 
phase (see Fig.1)\cite{dattagupta}.
We will be interested in analyzing questions such as what is the analogue of 
`phase ordering' in 
classical systems (see \cite{dattagupta}) when the quantum mechanical system of Fig.1 is subjected to a sudden 'quenching' along the $\Gamma$-axis across $\Gamma_{c}$, maintaining the temperature at $T=0$. Because a change in the value 
of $\Gamma$ (from above $\Gamma_{c}$ to below) will inevitably lead to an irreversible transition from one equilibrium configuration to another, dissipation needs to be dovetailed to the discussion. Again, a straightforward and widely studied model of quantum dissipation, popularized in recent years by Ford et al \cite{ford}, and
Caldeira and Leggett\cite{caldeira-leggett}, is the one in which the quantum 
system at hand is linearly coupled to a bosonic bath.
TIM is an effective model which has application from solid state materials of
rare earth magnets to
a coupled Josephson arrays\cite{dattagupta-book}, where dissipation arises in a 
natural way. 
Our model Hamiltonian can then be written,
\begin{eqnarray}
H &=&- \frac{1}{2}\sum_{i\neq j}J_{ij} s_{i}^{z}s_{j}^{z} - \Gamma \sum_{i} s_{i}^{x}\nonumber\\
&+& \frac{1}{\sqrt{N}}\sum_{i,k} s_{i}^{x}g_{k}(b_{k}^{\dagger} + b_{k}) 
+ \sum_{k} \hbar \omega_{k} b_{k}^{\dagger} b_{k},
\label{ham-ising}
\end{eqnarray}
where $N$ is the total number of spins and $g_{k}$ is a coupling constant. 
When $g_{k}=0$ and the Ising coupling $J_{ij}$ is treated in the mean field 
approximation(MFA), we obtain the phase diagram, depicted in Fig.1.

\begin{figure}
\rotatebox{0}{
\includegraphics*[width=8cm]{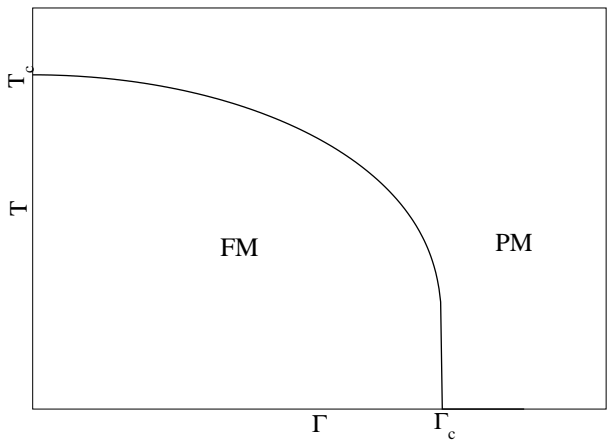}}
\caption{Mean field phase diagram of a TIM in which the abscissa represents the transverse field $\Gamma$ whereas the ordinate depicts the temperature $T$. When $\Gamma = 0$, we have the usual Curie transition at $T_{c}$. On the other hand at $T = 0$, there is a QPT at $\Gamma_{c}$ from a ferromagnetic(FM) to a 
paramagnetic(PM) phase.}
\label{fig1}
\end{figure}

Several limiting cases of Eq(\ref{ham-ising}) have received attention in recent 
literature. If $J_{ij} = 0$ and the term $s_{i}^{x}$ in the coupling with the 
heat bath is
replaced by $s_{i}^{z}$, Eq(\ref{ham-ising}) yields the celebrated spin-boson model of quantum dissipation \cite{leggett2}. When the Ising interaction is replaced by a Zeeman coupling with an
external field $h$ on a single ($N=1$) spin, the model in Eq(\ref{ham-ising}) describes low-temperature dissipative quantum tunneling in an asymmetric double well \cite{grabert,weiss-wollensak,dattagupta-grabert}. Additionally, if $h=0$, one has a symmetric double-well, tunneling in which can be impeded, leading 
to localization, when the coupling with the 
bosonic bath exceeds a certain critical value\cite{chakravarty,bray-moore}. This case is also relevant for a Kondo impurity of spin one-half (described by $\vec{s}$), in interaction with a conduction electron bath, which can be modeled in terms of bosons as far as the electron-hole excitations near the Fermi 
surface 
are concerned\cite{chang-chakravarty}.
For $\Gamma = 0$, this model can be viewed as an generalization of the Dicke 
model of `superradiance'
where the Ising term includes an additional atom-atom interaction\cite{dicke}. 

One other important application of 
Eq(\ref{ham-ising})
ensues in the case wherein the range of the Ising interaction is infinite, i.e,
\begin{equation}
J_{ij} = \frac{J}{N},
\label{coupling}
\end{equation}
a constant. This situation is the one in which the MFA to the Ising model (in the absence of the transverse field $\Gamma$) becomes exact and will occupy much of our attention below. Writing the total spin components as
\begin{equation}
S_{z} = \sum_{i} s_{i}^{z},~~~
S_{x} = \sum_{i} s_{i}^{x},
\label{totalspin}
\end{equation}
we have from Eq(\ref{ham-ising}) and Eq(\ref{coupling}),
\begin{equation}
H = -\frac{J}{2N} S_{z}^{2} - \Gamma S_{x} + \sum_{k} \frac{g_{k}}{\sqrt{N}} S_{x} (b_{k} + b_{k}^{\dagger}) + \sum_{k} \hbar \omega_{k}b_{k}^{\dagger}b_{k}.
\label{ham-largespin}
\end{equation}
If we leave aside the coupling term $g_{k}$, the spin-Hamiltonian in 
Eq(\ref{ham-largespin})
represents a `Molecular Magnet' characterized by a single-ion anisotropy energy $J/N$, and subjected further to a transverse field $\Gamma$\cite{barbara}. 
This problem has been investigated in great detail in recent years in the context of 'macroscopic magnetization tunneling' when the value of $S$ can be pretty large such as $S=10$. The additional coupling to the bosonic bath when $g_{k}$ is switched on, enables us to treat the effect of dissipation on this tunneling behavior\cite{palacios}.
The preceding remarks then underscore the versatility and relevance of the model Hamiltonian $H$ in Eq(\ref{ham-ising}) in a variety of applications to current topics in the condensed matter. In the sequel we shall analyze diverse mean-field and low-temerature properties of $H$, keeping the underlying QPT in mind.
With this background to the formalisms developed here, the paper is organized as follows.

In section II we first carry out a unitary transformation on the Hamiltonian in Eq(\ref{ham-ising}), that has been borrowed from the literature on polaron-physics\cite{dattagupta,holstein,silbey}. This transformation enables the original coupling constant $g_{k}$(proportional to a variational parameter $f_{k}$) 
to be elevated to an exponential function, thus facilitating an analysis that works even in the regime of strong coupling to the 
bath. The transformed Hamiltonian, rewritten in terms of $f_{k}$, is then 
treated in section III in the MFA to evaluate the associated density matrix from 
which the Helmholtz free-energy and an equation of state for the magnetization
 can be calculated. In section IV we focus our attention to spin wave like excitations near the absolute zero of temerature. The section V is devoted to a novel aspect of entanglement, important in the contemporary issue of quantum information process, in the context of two coupled Ising spins. Finally in section
VI we present a brief summary. 

\section{Effective Spin-Hamiltonian}
In this section we derive an effective Hamiltonian of the spin system, starting from 
Eq(\ref{ham-ising}).
For this, we integrate out the 
bosonic degrees of freedom of the heat bath. Here we adopt the usual definition of the 'effective partition function' $Z_{eff}$ of the spin system\cite{weiss},
\begin{equation}
Z_{eff} = Tr_{S+B} e^{-\beta(H_{s} + H_{B} + H_{I})}/Tr_{B}e^{-\beta H_{B}}.
\end{equation}
where $H_s$ is the Hamiltonian of the 'transverse field Ising model', $H_{B}$ describes the non-interacting bosonic degrees of the heat bath and $H_{I}$ is the interaction term with the heat bath. To decouple the spin-system from the bosonic heat bath, we subject the original Hamiltonian to a unitary transformation,
\begin{equation}
\tilde{H} = UHU^{-1},
\end{equation}
where,
\begin{equation}
U = exp\left[-\sum_{i}s_{i}^{x}\sum_{k}\frac{f_{k}}{\sqrt{N}}(b_{k} - 
b^{\dagger}_{k})\right].
\label{eq-unitary}
\end{equation}

At this stage we treat $f_{k}$ as a variational parameter which can be 
determined from the minimization of the free-energy of the total system. As a special case, if we consider a non-interacting spin system by setting 
$J_{ij} = 0$, we notice that the total Hamiltonian can be diagonalized by the 
unitary transformation given in Eq.(\ref{eq-unitary}) with $f_{k} = g_{k}/\hbar \omega_{k}$.
Motivated by this observation, the above mentioned variational method has been applied to a 
single-spin($N=1$) `spin-boson' model, which successfully captures the 'Kondo' like localization transition\cite{silbey,turlakov}.

Following the unitary transformation in Eq.(\ref{eq-unitary}) on Eq.(\ref{ham-largespin}),
\begin{eqnarray}
\tilde{H}  = & & -\frac{J}{2N}[-\frac{N}{4} + S_{z}^{2} + 
\frac{1}{2}(S_{z}^{2}- S_{y}^{2})(\cosh(-2\hat{F}) - 1) \nonumber\\
& &-\frac{\i}{2}(S_{z}S_{y} + S_{y}S_{z})
\sinh(-2\hat{F}) ] -\Gamma S_{x}\nonumber\\  
& & + \frac{S_{x}}{\sqrt{N}} \sum_{k}(b_{k} + b^{\dagger}
_{k})(g_{k} - \hbar \omega_{k} f_{k})\nonumber\\ 
& & -\frac{S_{x}^{2}}{N}\sum_{k}(2 f_{k} g_{k}
- \hbar\omega_{k} f_{k}^{2}) + \sum_{k} \hbar \omega_{k} b_{k}
^{\dagger}b_{k}
\label{transformed-hamil}
\end{eqnarray}
where, $ \hat{F} = \sum_{k}\frac{f_{k}}{\sqrt{N}}(b_{k} - b_{k}^{\dagger})$,
and $S_{x,z}$ are components of the total spin as given in 
Eq.(\ref{totalspin}).
Now we approximate the total density-matrix as a direct product form 
$\rho = \rho_{s} \bigotimes
 \rho_{B}$, where $\rho_{s}$ and $\rho_{B}$ denote the density matrices of 
the spin system and free bosons respectively.
After integrating out the bosonic modes of the system, we obtain the effective Hamiltonian $H_{eff}$ of the spin system:
\begin{equation}
Tr_{B}\tilde{H} = H_{eff} + \frac{J}{8} + F_{B},
\end{equation}
where $F_{B}$ is the free energy of noninteracting bosons and,
\begin{equation}
H_{eff}=-K_{z}S_{z}^2 -K_{y}S_{y}^{2}-K_{x} S_{x}^{2} -\Gamma S_{x}.
\label{heisenberg-hamil}
\end{equation}

The Hamiltonian above describes a fully anisotropic Heisenberg-model in the 
presence of a magnetic field 
along the x-axis, when it is written using Eq.(\ref{totalspin}). Effective 
coupling strengths are given by,
\begin{eqnarray}
K_{x} &=& \frac{1}{N} \sum_{k} (2 f_{k}g_{k} - \hbar \omega_{k} f_{k}^{2}),
\nonumber\\
K_{y} &=& \frac{J}{4N}\left[1 - exp\{-\frac{2}{N}\sum_{k} f_{k}^{2} 
\coth(\hbar \omega_{k}/2k_{B}T)\}\right],\nonumber\\
K_{z} &=& \frac{J}{4N}\left[1 + exp\{-\frac{2}{N}\sum_{k} f_{k}^{2} \coth(\hbar \omega_{k}/2k_{B}T)\}\right].
\label{effective coupling}
\end{eqnarray}
Going back to the notation of total spins (see Eq(\ref{totalspin})) we see that the terms proportional to $K_{x}$ and $K_{y}$ generate new physics in the context of molecular magnets. 
From minimization of the free-energy of the total system with respect to the variational parameter $f_{k}$, we obtain,
\begin{equation}
f_{k} = \frac{2g_{k}\langle S_{x}^{2}\rangle/N}{[2\hbar\omega_{k}\frac{
\langle S_{x}^{2}\rangle}{N} + \frac{J\tilde{K}}{N^{2}}(\langle S_{z}^{2}
\rangle - \langle S_{y}^{2}\rangle )\coth(\hbar \omega_{k}/2k_{B}T)]},
\label{selfcons_f}
\end{equation}
where 
\begin{equation}
\tilde{K} = exp\left[-\frac{2}{N}\sum_{k}f_{k}^{2} \coth(\hbar \omega_{k}/
2k_{B}T)\right].
\label{selfcons_k}
\end{equation}
Thermodynamics of the spin-system described by the effective Hamiltonian Eq.(\ref{heisenberg-hamil})can be obtained by solving the self-consitent equations 
described in Eq(\ref{selfcons_f}) and Eq(\ref{selfcons_k}), as described below.

\section{Phase transition of quantum Ising model: $N\rightarrow \infty$}
In this section we consider the case wherein the number of spins $N\rightarrow 
\infty$, which corresponds to an `infinite range' quantum ising model in the 
thermodynamic limit. For the corresponding classical model it is known that 
the MFA is exact in this limit.
\subsection{Mean-Field approximation}
Within the MFA the phase diagram of the above model can be sloved analytically. Here we assume that the density matrix of the spin system can be written as product of the density matrix of each spins $\rho_{s} = \prod_{i} \bigotimes \rho_{i}$. The density matrix of each spin can further be expressed as,
\begin{equation}
\rho_{i} = \frac{1}{2}(1 + m \sigma_{z} + \chi \sigma_{x}),
\end{equation}     
where the order parameters are $\langle s_{i}^{z}\rangle = m/2$ and $\langle s_{i}^{x}\rangle = \chi/2$.
From the selfconsistency equation for $f_{k}$ Eq(\ref{selfcons_f}) we obtain,
\begin{equation}
f_{k} = \chi^{2} g_{k}\left[\frac{Jm^{2} \tilde{K}}{2N} \coth(\hbar 
\omega_{k}/2k_{B}T) + \chi^{2}\hbar \omega_{k}\right]^{-1}.
\end{equation}
If we assume $\frac{1}{N}\sum_{k}g_{k}^{2}/\hbar^2 \omega_{k}^2 \rightarrow 0$, in the limit of $N\rightarrow \infty$, we have $\tilde{K} \rightarrow 1$.

The free energy of the system is given by,
\begin{eqnarray}
&&F = -\frac{J N m^{2}}{8}
-\Gamma \frac{N \chi}{2} -\frac{N\chi^{2}}{4} 
\sum_{k} (2 f_{k}g_{k} - \hbar \omega_{k} f_{k}^{2}) \nonumber\\
&&+ k_{B}T N\left[\frac{1}{2}(1 + \xi)\ln(1 + \xi) + \frac{1}{2}(1 - \xi)\ln(1 - \xi)
-\ln2 \right] \nonumber\\
&&+ F_{B}
\label{MFA_spin}
\end{eqnarray}
where $\xi = \sqrt{m^2 + \chi^2}$ and $F_{B}$ is the free energy of the 
bosons.
Minimizing the free energy per particle $F/N$ with respect to the parameters 
$m$ and $\chi$ we obtain:
\begin{eqnarray}
-\frac{Jm}{4} + \frac{kT m}{2\xi}\ln\left[\frac{1 + \xi}{1 - \xi}\right] = 0
\\
-\frac{\Gamma}{2} - \frac{\chi \lambda }{2} +  \frac{kT \chi}{2\xi}\ln
\left[\frac{1 + \xi}{1 - \xi}\right] = 0,
\label{MFA_eqn}
\end{eqnarray}
where $\lambda = \sum_{k}g_{k}^2/\hbar \omega_{k}$. The ferromagnetic phase is defined by $m \neq 0$ and the phase boundary can be obtained from
\begin{equation}
\Gamma/(J/2 - \lambda) = \tanh\left[\frac{J\Gamma}{4k_{B}T(J/2 - \lambda)}\right].
\label{MFA-magnetisation}
\end{equation}
A three-dimensional plot of the phase diagram is shown in Fig. 2. As expected 
of course, for $\lambda = 0$, Eq. (\ref{MFA-magnetisation}) yields the usual 
equation of state and the phase diagram depicted earlier in Fig. 1. The latter 
also bears out the expectation that as the strength of the coupling  to the 
heat bath $\lambda$ increases, the QCP on the $\Gamma$-axis is suppressed.

\begin{figure}
\rotatebox{0}{
\includegraphics*[width=\linewidth]{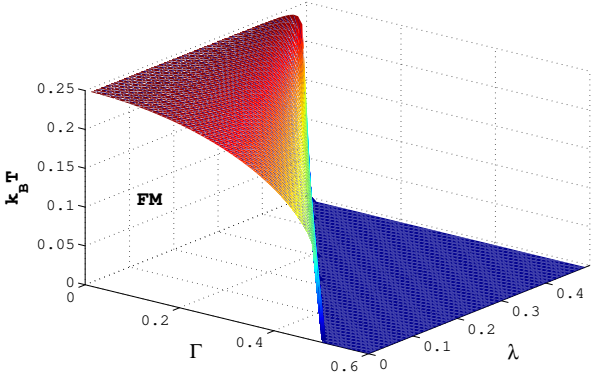}}
\caption{Phase boundary of ferromagnetic phase of a TIM coupled to a heat bath (see Eq.(\ref{MFA-magnetisation})). Temperature $T$, transverse field $\Gamma$ and strength of dissipation $\lambda$ are measured in units of $J$. The region under the curve represents the ferromagnetic(FM) phase.}
\label{fig2}
\end{figure}

\subsection{Semiclassical analysis}
In this subsection we calculate the effective partition function of the spin-system semi-classically and establish the validity of the variational method and MFA for a large spin system ($N\rightarrow \infty$). We can decompose a direct product of $N$ spin $1/2$ systems into a direct sum of total spin $S$, where the magnitude of the spin varies from $N/2$ to $0$ or $1/2$ (depending on even or odd $N$). In terms of the total spin operator $S$, the partition function can be written as\cite{lieb}
\begin{eqnarray}
&&Z = \sum_{\{s_{i}\}} Tr_{B} e^{-\beta H}
=\sum_{S=0,1/2} ^{N/2} P(S)Tr_{S} Tr_{B} \nonumber\\
&&e^{-\beta\left[-\frac{J}{2 N} S_{z}^2 - \Gamma S_{x} + \frac{1}{\sqrt{N}}S_{x}\sum_{k} g_{k} (b_{k} + b_{k}^{\dagger}) + \sum_{k} \hbar \omega_{k} b_{k}^{\dagger} b_{k} \right]},
\label{z_s}
\end{eqnarray}
where $P(S)$ is the number of ways the total spin $S$ can be formed\cite{lieb} and is given by
\begin{equation}
P(S) =  \frac{(2S + 1)N!}{(N/2 -S)! (N/2+S+1)!}.
\end{equation}

For large $S$ we can calculate the partition function of spin systems 
classically\cite{lieb} and replace the trace by the integral,
\begin{equation}
Tr_{S}
\rightarrow \frac{2S + 1}{4 \pi} \int \sin\theta d\theta d\phi 
\end{equation}
where the components of the total spin $\vec{S}$ are given by $(S\sin\theta \cos\phi,S\sin\theta\cos\phi,S\cos\theta)$.
Using the coherent-states for bosons we integrate out the bosonic degrees of freedom and obtain the effective partition function of the spin system as,
\begin{eqnarray}
&&Z_{eff}=\sum_{S = 0,1/2}^{N/2} P(S) \frac{(2S + 1)}{4\pi}\int\sin\theta d\theta d\phi \nonumber\\
&&e^{-\beta\left[ - \frac{J}{2N} S^2 \cos^{2} \theta - \Gamma
 S \sin \theta \cos \phi - \frac{\lambda}{N} S^{2} \sin^{2} \theta \cos^{2} \phi \right]}
\end{eqnarray} 
The minimum value of the energy is obtained for $\phi = 0$. For large $S$ we introduce two 
variables $r = \frac{S}{N/2}$ and $x = \cos\theta$. In terms of these two variables we can write the effective partition function as:
\begin{equation}
Z_{eff} = N_{r} \int_{0}^{1} dr dx e^{-\beta N f(r, x)},
\end{equation}
where $f(r,x)$ is the free energy per particle and $N_{r}$ is the a constant.
Here we have used Stirling's approximation,
\begin{equation}
lnP(S) \approx -\frac{N}{2}\left[ (1 + r)ln(1 + r)
+ (1 - r)ln(1- r) - 2ln2\right],
\end{equation}
which is the entropy in MFA.
The free energy per particle is given by:
\begin{eqnarray}
&&f(r,x)=-\frac{J}{8} r^{2} x^{2} - \frac{\Gamma r}{2} \sqrt{1 - x^{2}} - \frac{\lambda}{4} r^{2}(1 - x^{2})+\nonumber\\
&&\frac{k_{B} T}{2}  \left[ (1 + r)ln(1 + r)+(1- r)ln(1- r)\right].
\end{eqnarray}

Within a saddle point approximation we minimize $f(r,x)$ and 
obtain the following set of equations:
\begin{eqnarray}
&&\frac{\partial f(r,x)}{\partial x} =  0, \nonumber \\
&& r \sqrt{1 - x^{2}} = \frac{\Gamma}{J/2 - \lambda}, \\
&&\frac{\partial f(r,x)}{\partial r} =  0, \nonumber\\
&&- \frac{J}{2} r x^{2} - \Gamma \sqrt{1 - x^{2}} - \lambda r (1 - x^{2}) \nonumber\\
&&+ k_{B}T \ln[\frac{1 + r}{1 - r}] = 0.
\label{saddlepoint}
\end{eqnarray}
Defining the magnetisation along z-axis as $m/2 = rx/2$, and using Eq(\ref{saddlepoint}) at the saddle point of the free energy function $f(r,x)$ we obtain,
\begin{equation}
\left[ m^{2} + \frac{\Gamma^2}{(J/2 - \lambda)^2}\right]^{1/2} = \tanh\frac{J}{4 k_{B} T}\sqrt{ \left[ m^{2} + \frac{\Gamma^2}{(J/2 - \lambda)^2}\right]},
\end{equation}
which checks with the earlier Eq.(\ref{MFA-magnetisation}).

\section{Collective excitations near $T=0$}
In this section we analyze the low temperature spin wave like excitations by 
calculating the low-lying energies of the spin system
coupled to a heat bath. In the previous section we used the total spin 
representation
$S^{a} =\sum_{i} s_{ia}$, where $a=x,y,z$, to evaluate the partition function 
semiclassically. Quantum effects become relevant near zero temperature and 
consequently, the
total spin takes a large value $S \sim N/2$, in order to minimize the free energy.
In the limit of large $S$, we can derive an effective Hamiltonian describing the 
quantum fluctuations at low temperatures.
In the Holstein-Primakoff representation, the components of total spin $S$ are
given by,
\begin{eqnarray}
S_{z}&=&S  - b^{\dagger}b,\\
S_{-}&=&b^{\dagger}\sqrt{2S - b^{\dagger}b},\\
S_{+}&=&\sqrt{2S - b^{\dagger}b} b.
\label{spin_hp}
\end{eqnarray}
where b($b^{\dagger}$) satisfy bosonic commutation rules.
We assume that the total spin vector $S$ makes an angle $\theta$ with
the z-axis and its projection on x-y plane makes an angle $\phi$ with the 
x-axis. 
Now we can write:
\begin{eqnarray}
S^{\prime z}&=&(S - b^{\dagger}b)\cos\theta -\sqrt{S/2}(b^{\dagger} + b)\sin\theta \\
S^{\prime x}&=&\sqrt{S/2}\left[(b^{\dagger} + b)\cos\theta \cos\phi + \imath 
(b - b^{\dagger})\sin\phi \right] \nonumber\\
&&+ (S -  b^{\dagger}b)\sin\theta \cos\phi
\label{rot-hp}
\end{eqnarray}

Substituting Eq(\ref{rot-hp}) in the Hamiltonian Eq.(\ref{ham-largespin}), we obtain,
\begin{eqnarray}
H&=&\frac{J}{8}-\frac{J}{2N}[(S -  b^{\dagger}b)^{2}\cos^{2}\theta
+\frac{S}{2}(b + b^{\dagger})^{2}\sin^{2}\theta \nonumber\\
&-&2\sqrt{\frac{S}{2}}(S -  b^{\dagger}b)
(b^{\dagger} + b)\cos\theta \sin\theta ]\nonumber\\
&-&[\sqrt{\frac{S}{2}}\{(b^{\dagger} + b)\cos\theta \cos\phi 
+\imath
(b - b^{\dagger})\sin\phi\}\nonumber\\ 
&+&(S -  b^{\dagger}b)\sin\theta \cos\phi ]
\times[\Gamma -\frac{1}{\sqrt{N}}\sum_{k} g_{k} (b_{k}^{\dagger} + b_{k})]\nonumber\\  
&+&\sum \hbar \omega_{k} 
b_{k}^{\dagger} b_{k} + O(1/S).
\label{ham-hp}
\end{eqnarray}
After performing a transformation $c_{k}^{\dagger} = b_{k}^{\dagger} + \frac{S}{\sqrt{N}\hbar\omega_{k}}g_{k}\sin\theta\cos\phi$, we obtain from the above Hamiltonian the classical energy of large spin $S$ (part of the Hamiltonian which is proportional to $N$ for large $S$),
\begin{equation}
H_{0} = -\frac{J S^{2}}{2N}\cos^{2}\theta - \Gamma S \sin\theta \cos\phi
-\frac{\lambda}{N} S^{2} \sin^{2}\theta\cos^{2}\phi
\end{equation}
Again, a minimum of $H_{0}$ can be found for $\phi = 0$ and
$\cos\theta = 0$ for the paramagnetic state. In the ferromagnetic state of spin $S$, the minimum energy
can be achieved for $\phi=0$ and,
\begin{equation}
\frac{2}{N}S(\frac{J}{2} - \lambda) \sin\theta = \Gamma.
\label{eq-theta}
\end{equation}
The above equation determing the minimum classical energy is equivalent to the  equation derived from the saddle point
approximation to the free energy described in sec.IIB (see Eq(\ref{saddlepoint})).
With this choice of $\theta$ and $\phi$, we notice that the terms linear in fluctuation
operators ($b$ and $c_{k}$) in the Hamiltonian (Eq(\ref{ham-hp})) vanish.

Finally the Hamiltonian describing the fluctuations can be exactly mapped on 
to the Caldeira-Leggett
model\cite{caldeira-leggett} describing an oscillator coupled to a heat bath,
\begin{eqnarray}
H_{2}&=&\epsilon b^{\dagger}b + \Delta (b^{\dagger} + b)^{2}\nonumber\\ 
&+&(b + b^{\dagger})\sum_{k} \tilde{g}_{k}(c_{k}^{\dagger} + c_{k}) + 
\sum_{k}\hbar\omega_{k}c_{k}^{\dagger}c_{k}
\label{ham-fluc}
\end{eqnarray}
where different parameters in the above Hamiltonian can be written in terms 
of the spin $S$ and its 
orientation $\theta$,
\begin{eqnarray}
\epsilon &=& \frac{JS}{N}\cos^{2}\theta + \frac{2\lambda S}{N}\sin^{2}\theta + \Gamma \sin\theta ,\\
\Delta &=& -\frac{JS}{4N}\sin^{2}\theta,\\
\tilde{g}_{k} &=& g_{k}\sqrt{\frac{S}{2N}}\cos\theta .
\end{eqnarray}
Following the work of Ambegaokar and Hakim\cite{hakim}, we can diagonalize 
the above Hamiltonian
quadratic in bosonic operators by means of the canonical transformations,
\begin{eqnarray}
c_{\alpha} &=& \sum_{\beta} A_{\alpha \beta} \tilde{c}_{\beta} + B_{\alpha 
\beta} 
\tilde{c}^{\dagger}_{\beta}, \nonumber\\
c^{\dagger}_{\alpha} &=& \sum_{\beta} B_{\alpha \beta} \tilde{c}_{\beta} + A_{\alpha \beta} 
\tilde{c}^{\dagger}_{\beta} ,\nonumber\\
\label{canonical-transform}
\end{eqnarray}
where we denote the original bosonic operators as $\{c_{\alpha}\} =\{b,c_{k}\}
$. The Hamiltonian can be written in diagonal form $H_{2} = \sum_{\beta} E_{\beta}\tilde{c}^{\dagger}_{\beta}\tilde{c}_{\beta}$ in terms of a new set of bosonic operators. The set of energies $E_{\beta}$ 
describes the low-lying excitation energies of the many-body system.
Equation of motion of the operators can be obtained from the Hamiltonian, 
yielding
\begin{eqnarray}
\imath\frac{db}{dt}&=&\epsilon b +2\Delta (b^{\dagger} +b)+\sum_{k}g_{k}(c^{
\dagger}_{k} + c_{k}),\\
\imath\frac{da_{k}}{dt}&=&\hbar\omega_{k}c_{k} +g_{k}(b^{\dagger} + b),\\
\imath\frac{d\tilde{c}_{\beta}}{dt}&=&E_{\beta}\tilde{c}_{\beta}.
\label{eq-motion}
\end{eqnarray}
Substituting Eq(\ref{canonical-transform})in Eq(\ref{eq-motion}), the matrix 
elelements of the canonical transformation are obtained as,
\begin{eqnarray}
E_{\beta}A_{0\beta}&=&\epsilon A_{0\beta} + 2\Delta (A_{0\beta}+B_{0\beta})+\sum_{k}g_{k}(A_{k\beta}+B_{k\beta}),\nonumber\\
-E_{\beta}B_{0\beta}&=&\epsilon B_{0\beta} + 2\Delta (A_{0\beta}+B_{0\beta})+
\sum_{k}g_{k}(A_{k\beta}+B_{k\beta}),\nonumber\\
E_{\beta}A_{k\beta}&=&\hbar\omega_{k}A_{k\beta}+g_{k}(A_{k\beta}+B_{k\beta}),
\nonumber\\
-E_{\beta}B_{k\beta}&=&\hbar\omega_{k}B_{k\beta}+g_{k}(A_{k\beta}+B_{k\beta}).
\label{matrix-coeff}
\end{eqnarray}

Finally, we obtain the equation determining the excitation energies $E_{\beta}$,
\begin{equation}
E_{\beta}^{2} - \epsilon^{2} - 4 \epsilon \Delta = 4\epsilon\sum_{k} 
\frac{\tilde{g}_{k}^{2}\hbar\omega_{k}}{E_{\beta}^{2} - \hbar^{2}\omega_{k}^{2}}.
\label{excitation-en}
\end{equation}
At $T=0$ the total spin is $S=N/2$, and we can define two phases according to
the orientation of the large spin. When $\Gamma > J/2 - \lambda$, 
all spins are aligned along the x-axis, hence $\cos\theta = 0$ solution 
minimizes the classical energy of the system. The QCP is given by,
\begin{equation}
\Gamma_{c} = J/2 - \lambda.
\end{equation}
In the paramagnetic phase the fluctuation operators of spin ($b^{\dagger}$,$b$) decouple from the modes of the bath in the leading order, since
\begin{equation}
\tilde{g}_{k} = g_{k}\cos\theta/2 =0
\end{equation}
and they are coupled in the higher order terms of the Hamiltonian which are 
supressed by a factor of $1/\sqrt{N}$. Quantum fluctuations of the paramagnet 
can then be described by an effective harmonic oscillator
\begin{equation}
H_{2} = \hbar\omega_{p}\tilde{c}^{\dagger}_{0}\tilde{c}_{0},
\end{equation}
with the excitation energy:
\begin{equation}
\hbar\omega_{p} = \sqrt{(\lambda + \Gamma)(\lambda + \Gamma - J/2)}.
\end{equation}
Here we note that although the fluctuations of spins are decoupled from the 
bath modes, the frequency $\omega_{p}$ depends on dissipation.
At the QCP, the excitation energy vanishes as $(\Gamma - \Gamma_{c})^{1/2}$, 
which is in accordance with mean-field behavior.

\section{Generalized spin-boson model for two interacting spins}
In this section we address the issue of quantum information and Schr$\ddot{o}$dinger cat like state by considering our model-Hamiltonian Eq.(\ref{ham-ising}) with
$N=2$, which represents two interacting spins attached to a heat bath.
This generalizes the usual spin-boson model\cite{leggett2} by including spin-spin  interaction terms.

Following the procedure of unitary transformation and then tracing out the 
bosonic degrees of freedom (as mentioned in section II), we obtain the 
following effective Hamiltonian for the total spin,
\begin{equation}
H=-K_{z}S_{z}^{2}-K_{y}S_{y}^{2}-K_{x}S_{x}^{2}-\Gamma S_{x},
\label{spin2}
\end{equation}
which follows directly from Eq.(\ref{heisenberg-hamil}).
Since at zero temerature, the triplet state plays an important role, we obtain
the following eigenvalues and eigenfunctions for the triplet state:
\begin{equation}
\epsilon_{\pm} = -(K_{z} + K_{y})/2 -K_{x}\pm \sqrt{(K_{z} - K_{y})^{2}/4 + 
\Gamma^2}
\end{equation}
\begin{equation}
|\pm> = \frac{\cos\theta_{\pm}}{\sqrt{2}}(|\uparrow\uparrow\rangle + |\downarrow\downarrow\rangle) + \frac{\sin\theta_{\pm}}{\sqrt{2}}(|\uparrow\downarrow\rangle + |\downarrow\uparrow\rangle),
\end{equation}
with $\tan\theta_{\pm}=-(\epsilon_{\pm} +K_{z} + K_{x})/\Gamma$, and,
\begin{eqnarray}
\epsilon_{2}&=&-K_{z}-K_{y},\\
|2\rangle &=& \frac{1}{\sqrt{2}}(|\uparrow\uparrow\rangle - |\downarrow
\downarrow\rangle).
\end{eqnarray}
The singlet state has zero energy. For a system without 
dissipation( when $K_{y} =0$, $K_{x}=0$) the ground state wave function is 
$|-\rangle$, with
\begin{equation}
\cos(2\theta) = \frac{K_{z}/2}{\sqrt{K_{z}^{2}/4 + \Gamma^{2}}}.
\end{equation}
Unlike the Ising system in the thermodynamic limit (for $N\rightarrow\infty$), 
this state does not have a net magnetization and $\langle S_{z}\rangle = 0$.
The magnetization along the x-axis increases smoothly with increasing the
magnetic field $\Gamma$ and finally for very large $\Gamma$, the state
$|-\rangle$ is factorized in two spin states directed along $x$ axis.
\begin{figure}
\rotatebox{0}{
\includegraphics*[width=\linewidth]{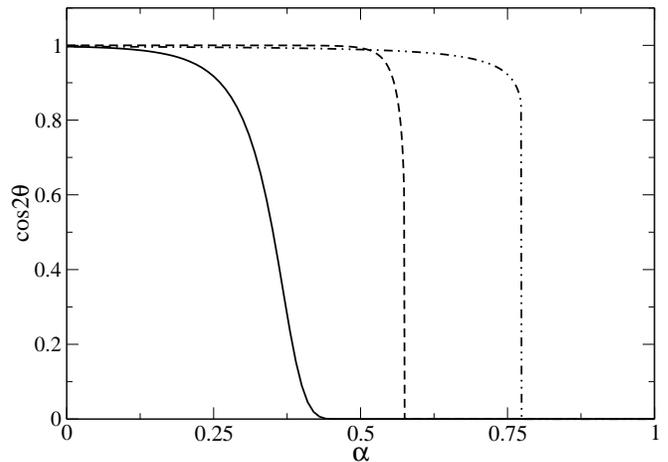}}
\caption{Variation of the parameter $\cos2\theta$ of the wavefunction $|-\rangle$ (which is also the `concurrence' of the wavefunction) with the change of the coupling $\alpha$ with the heat bath, for
$\Gamma/J =10^{-2}$, $\Omega_{c}/J=10^2$ (solid line), $\Gamma/J =10^{-4}$, $\Omega_{c}/J=10^2$ (dashed line) and $\Gamma/J =10^{-4}$, $\Omega_{c}/J=1$ (dot-dashed line).}
\label{fig3}
\end{figure}

Now switching on the coupling with the heat bath, we obtain,
\begin{equation}
\cos(2\theta) = \frac{J\tilde{K}/8}{\sqrt{(J\tilde{K}/8)^{2} + \Gamma^{2}}}.
\end{equation}
For the state $|-\rangle$,$\langle S_{x}^{2}\rangle = 1$,$\langle S_{z}^{2}\rangle = \cos^{2}\theta$,$\langle S_{y}^{2}\rangle = \sin^{2}\theta$, and hence, $f_{k} = g_{k}/[\hbar\omega_{k} + J\tilde{K}\cos(2\theta)/4]$.
Further, for an Ohmic heat bath, the spectral density is given by,
\begin{equation}
\sum_{k}g_{k}^{2}\delta(\omega -\omega_{k}) = \alpha \omega \theta(\Omega -\omega),
\end{equation}
where $\alpha$ is the coupling strength and $\Omega$ is the cut-off frequency
of the bath.
The renormalization factor $\tilde{K}$ can be obtained from the self-consistent
equation,
\begin{equation}
\tilde{K} = (\frac{J\tilde{K}\cos(2\theta)/4}{\Omega + J\tilde{K}\cos(2\theta)/4})^{\alpha} \exp\left[\frac{\alpha \Omega}{\Omega + J\tilde{K}\cos(2\theta)
/4}\right].
\end{equation}
Like in the spin-boson model, the parameter $\tilde{K}$ shows a crossover
behavior 
below $\alpha \approx 1$. Two spins become parallel to x-direction and the 
magnetization along x-axis sharply becomes $1$ when $\alpha$ crosses a 
critical value $\alpha_{c}$ for finite field $\Gamma$.

Also we focus on another aspect of this wavefunction. For $\theta =0$, the
wave-function $|-\rangle$ is maximally entangled as in a `cat-state'\cite{catstate}. In the other
limit $\theta = \pi/4$ (in the limit of large field $\Gamma$), the ground state
$|-\rangle$ can be factorized. The amount of entanglement between 
the two spins can be quantitatively calculated from the `concurrence' of 
the wave function of two spins, which is $\cos(2\theta)$ \cite{wooters}. It is interesting to notice
that the concurrence shows a crossover at $T=0$ by tuning the 
dissipative coupling strength $\alpha$ as shown in Fig.3.
 
\section{Summary and Outlook} 
In this paper we analyze the effect of dissipation on a model quantum 
system which undergoes a QPT. In various limiting cases this model represents 
different physical systems, starting from a TIM in the thermodynamic limit
(when the number of spins $N \rightarrow \infty$) to a generalized spin-boson
model of two interacting spins (for $N = 2$). Also this model can be viewed 
as an interacting version of a multi-mode Dicke model describing the atom-photon
interaction. In the context of a large spin our model has application to the 
area of nanomagnets. We have developed a self-consistent variational method to 
study the system for any number of spins, and the spin system can be described by an effective anisotropic 
Heisenberg model after integrating out the bosonic modes. The effective 
Heisenberg model shows two interesting effects, the renormalization of the 
orginal coupling between the z-component of spins similar
to the spin-boson model and generation of extra terms which couple the 
x and y-components and of spins. In the thermodynamic limit ($N\rightarrow 
\infty$) the model represents a TIM in presence of dissipation and the phase diagram
has been obtained within a variational-MFA which is in accordance with the 
saddle point approximation. It is interesting to note that for large $N$, a 
new coupling between the x component of spins plays an important role and 
destroys
the ferromagnetic phase for a smaller value of transverse field.
The QPT of the model at zero temperature has been studied using the 
Holstein-Primakoff transformation. The excitation energy vanishes as 
$(\Gamma - \Gamma_c)^{1/2}$ at the quantum critical point.
 
In the other limit, for $N=2$, the model generalizes the usual spin-boson 
model by including a spin-spin interaction along the z axis. Unlike a TIM, 
this two-spin system shows zero magnetization along the z-axis but displays a 
sharp change in magnetization along the x-axis. For small $N$, strong renormalization of spin-spin coupling of the original Hamiltonian becomes important
for the crossover phenomenon similar to the original spin-boson model.
From the point of view of quantum information theory the ground state of 
two-spins shows a transition from a maximally entangled `cat-state' to a 
factorizable
state where both spins are aligned along the x-axis.

In conclusion we have studied the effect of dissipation on a simple model of 
an interacting spin system which undergoes QPT. Further extension of the 
model for analyzing the 
effect of dissipation in dynamical quenching across a quantum critical point
that is also 
relevant for the dynamics and thermodynamics of nanomagnets, as alluded to in 
sec. I, is left for future work\cite{SDA}.

\end{document}